\documentclass[ floatfix,reprint,amsmath,amssymb,aps,prb]{revtex4-1}
\usepackage{xcolor}
\usepackage{bm,hyperref}
\usepackage{textcomp}
\usepackage{amsmath}
\usepackage{amssymb}
\usepackage{graphicx}
\usepackage{esint}
\usepackage{color}
\usepackage{float}
\usepackage{enumitem}

\begin{document}

\title{Perturbing beyond the shallow amplitude regime: Green's function scattering formalism with Bloch modes}

\author{A. Abass$^{1,*}$, A. Martins$^2$, S. Nanz$^3$, B.-H.V. Borges$^2$, E. R. Martins$^2$, C. Rockstuhl$^{1,3}$ }
\affiliation{\mbox{$^1$Institute of Nanotechnology, Karlsruhe Institute of Technology (KIT), 76021 Karlsruhe, Germany} 
\mbox{$^2$Department of Electrical and Computer Engineering, University of S\~{a}o Paulo, CEP 13566-590, São Carlos, SP, Brazil}
\mbox{$^3$Institute of Theoretical Solid State Physics, Karlsruhe Institute of Technology, Wolfgang-Gaede-Str. 1, 76131 Karlsruhe, Germany}
$^*$Corresponding author: aimi.abass@kit.edu}

\begin{abstract}

We present a Bloch modes' based Green's function scattering formalism for cost efficient forward modelling of disordered binary surface textures. The usage of Bloch modes of an unperturbed reference ordered system as ansatz allows our formalism to address surface scattering beyond the shallow amplitude regime.  The main advantage of our formalism is the possibility to utilize a small amount of plane waves to represent the assumed Bloch modes thereby reducing computational costs, while still allowing one to estimate the scattering response to all channels accessible by the disordered system. Benchmarking calculations discussed in the paper demonstrate how the usage of our Bloch modes ansatz provides an excellent estimate of the scattering response over an important regime of disorder. As an example of our method's strength, we examine an electrically decoupled binary light trapping texture and demonstrate how introducing disorder may improve light incoupling into the considered solar cell structure.

\end{abstract}
\maketitle

\section{Introduction}

Having on-demand large-area interfaces with desired light interaction is a long-standing vision in the field of optics. Such interfaces have a plethora of applications for various optoelectronic devices, such as solar cells or OLEDs \cite{Abass2013,Gomard2016,Nanz2018}. For many of these applications, not only that the interfaces should operate optimally in providing a desired scattering response, they should also operate either over an extended spectral domain or  with a wide acceptance angle. Such performance demand, however, forces one to consider instead quasi-periodic or disordered structures\cite{Abass2012,Oskooi2012,Martins2013,Brongersma2014,Piechulla2018}. 

Computationally addressing and optimizing disordered or aperiodic textures typically requires huge computational efforts due to the necessity of accommodating extended computational domains. A rigorous numerical treatment of optical wave scattering by disordered structures typically requires a large memory space and/or a long computational time. To tackle this challenge, extensive efforts have been dedicated to develop approximate and computationally efficient analytical and semi-analytical models to address light scattering by complex interfaces \cite{Elfouhaily:2004,Schroder:11,Gonzalez-Alcalde:16}. 

There have been methods based on the Kirchhoff approximation which assumes a surface textures with a slowly varying slope typically valid only when the
correlation length of the surface is much larger than the
wavelength and when small angles of incidence are considered \cite{Thorsos:1988,WangKirchoff:2005}. The generalized Harvey-Shack scattering formalism offers a non-paraxial treatment to the scattering problem and can addresses moderately rough surface textures. The Harvey-Shack formalism is, however, a scalar diffraction theory, which considers surface texture solely in terms of macroscopic statistical quantities \cite{Schroder:11,Krywonos:11}. 

Methods based on perturbation theory (Rayleigh-Rice approximation-based theories) have been shown to offer a full vectorial treatment \cite{Rice:1951,Saillard:2001,Soubret2001}. They can treat small correlation lengths as well as large incidence and scattering angles but they are typically limited to surface texture amplitudes much smaller than the wavelength \cite{Elfouhaily:2004}. Various Green's function formulations have been proposed and discussed in the literature to address the scattering of rough surfaces, but they have limited surface amplitude validity range as with Rayleigh-Rice based methods  \cite{Sipe:87,Paddon1998,Guerin:04,Rottenberg:2011}. 

Despite the limited validity range, perturbation theory based methods can capture important photonic effects such as the excitation of surface plasmon polaritons (SPP) \cite{Rottenberg:2011,Reitich:13}. Novel scattering phenomena, such as the Yoneda and the Brewster scattering effects, can already be accounted for with first order perturbation considerations\cite{Banon2010}. Scattering calculations based on the Born approximation extended with the modified  Fraunhofer scattering approach, which better accounts for the phase shift along the surface corugation, provides fairly reliable predictions for relevant thin-film solar cell structures\cite{Jager2009,Domine2010}. The validity regime concerning the possible surface texture amplitude to be captured with current perturbation-based approaches can be limitedly extended by considering higher order perturbation terms at the cost of severe increase in complexity \cite{Soubret2001,Demir2003,Guerin:04,Demir2012,Nicholls:15}.    

In a recent publication, we developed an alternative fully vectorial Green's function-based scattering formalism for forward and inverse modelling of aperiodic surface textures \cite{Abass2017}. With our approach, we can analytically formulate the inverse problem of designing a multiresonant quasiperiodic scattering system in terms of multivariate coupled polynomial equations of the surface texture Fourier amplitudes. The resulting polynomial equation system is in turn computationally more efficient to solve.  A recently proposed vectorial Green's function-based method shows how one can address the scattering due to large surface amplitudes with relative computational efficiency by approximating the Green's function in terms of a limited number of resonant contributions\cite{Fehrembach2018}. The complexity and computational costs of these approaches, however, still tend to skyrocket when the number of resonant interactions is increased. 

Here, we introduce a physically intuitive Green's function perturbative scattering formalism that exploits Bloch modes as solution ansatz. Unlike Rigorous Coupled Wave Analysis (RCWA)\cite{Li:96,Li:97}, our formalism decouples the utilized modal field ansatz in the scattering region from the actual considered geometry.  The method allows flexible control of the accuracy and computational cost trade-off by the choice of the reference structure from which the Bloch Modes ansatz is taken. With that in mind, one can perform numerically efficient fully vectorial calculations of extended quasi-periodic or disordered systems beyond the shallow grating approximation. We demonstrate the application of the method by modelling disordered light trapping binary textures. 

Section II describes the usage of Bloch modes ansatz to solve the implicit integral equations in a Green's function scattering formalism. Section III describes how the scattering response is deduced through our perturbative  approach. Section IV discusses the numerical benchmarking of our method against Rigorous Coupled Wave Analysis (RCWA) in handling super-periodic and disordered systems. Section V discusses an example application of our method in considering light trapping structures in a c-Si solar cell system. A summary of the work is provided in section VI.
\begin{widetext}
\section{Unperturbed Bloch modes scattering formalism}

Here, we describe the usage of Bloch modes as ansatz in a Green's function k-space scattering formalism, which we refer to as Green's Method of Unperturbed Bloch Modes (GMoUB). We derive how  the implicit integral equation involved in the Green's function formalism is changed into a complete system of linear equations. 

\begin{figure}[h]
	\begin{center}
	\includegraphics[scale=0.44]{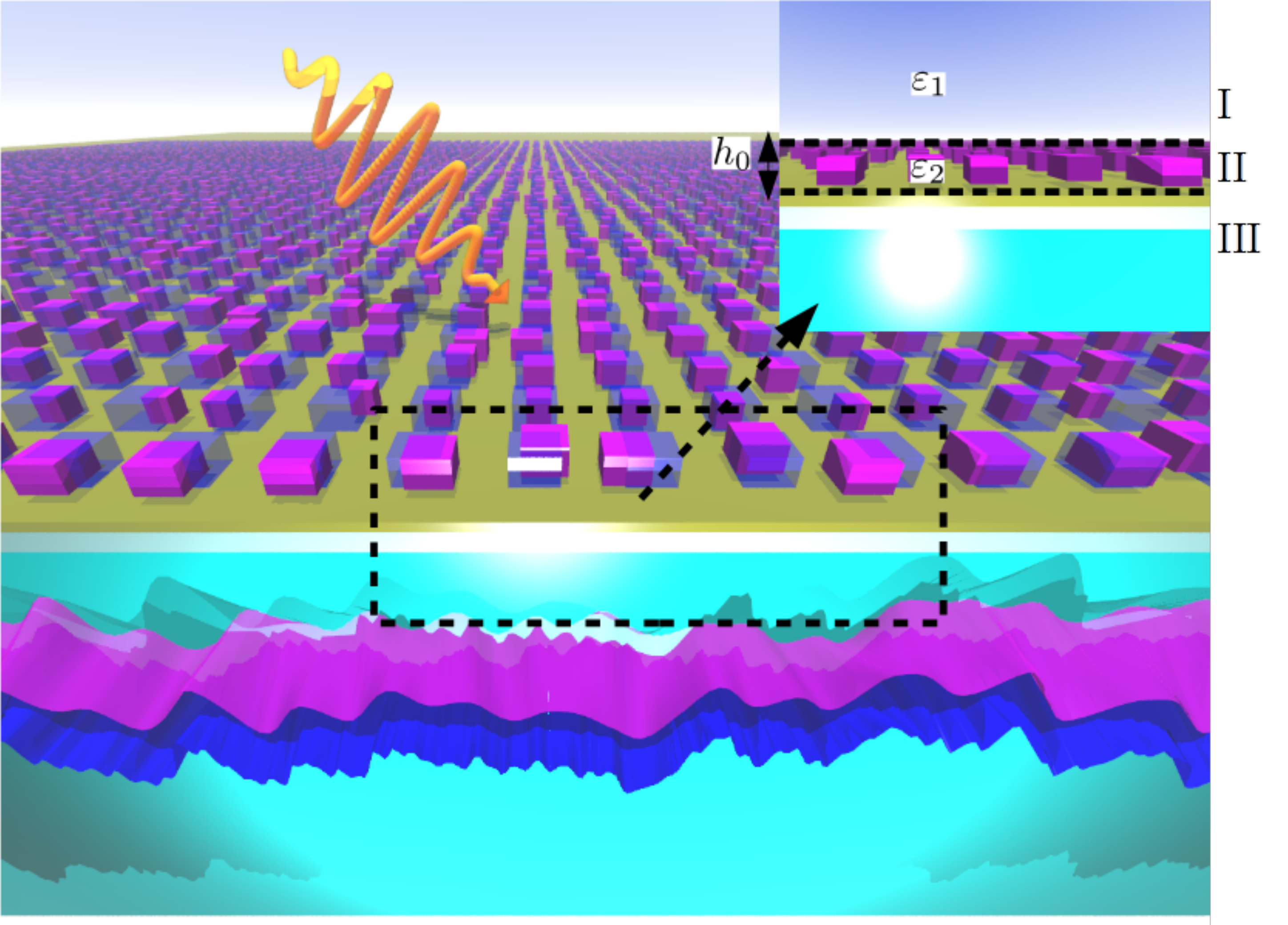}
	\caption{\textit{A sketch of the considered system and the proposed theoretical approach. Instead of rigorously considering the whole disordered structure  (purple array), we consider the modes in the scattering region to be that of a simpler reference periodic structure (blue periodic array)  to calculate the scattering response and thus obtain an approximate solution (bottom blue wavefront) to the actual response (purple wavefront).}}\label{fig1}
	\end{center}
\end{figure}

The system we consider is depicted in Fig.~\ref{fig1} where a disordered/quasiperiodic scattering structure is placed at the interface between two regions of different permittivities. We divide the system into three different regions. Region I is the medium of incoming wave, which we consider as a homogeneous halfspace with permittivity $\varepsilon_1$. Region II contains the scattering structure which may as well possess a different permittivity $\varepsilon_2$. Region III either consists of a semi-infinite substrate or a multilayer structure. 

As described in our previous work\cite{Abass2017}, the key to solving the scattering problem, whether forward or inverse, is in deducing the field in region II. Utilizing the Green's function formulation of scattering at an interface texture developed by Sipe\citep{Sipe:87,Rottenberg:2011}, the field region II can be described by 

\begin{align}
\mathbf{E}^{\varrho}_{\mathrm{II}}(\boldsymbol{\kappa}_{n},z)= & \mathbf{E}^{-,\varrho}_{\mathrm{II,inc}}(\mathbf{\kappa_{n}}) e^{-iw_{1,n}z}\nonumber +\mathbf{E}^{+,\varrho}_{\mathrm{II,inc}}(\boldsymbol{\kappa}_n) e^{iw_{1,n}z}\nonumber -\frac{\hat{\boldsymbol{z}}\otimes\hat{\boldsymbol{z}}}{\epsilon_{0}\epsilon_{r,1}}\mathbf{P}(\kappa_{n},z) \\
 & +\frac{ik_{0}^{2}}{2\varepsilon_{0}w_{1,n}}\left(\hat{\boldsymbol{\varrho}}_{1+}^{n}\otimes\hat{\boldsymbol{\varrho}}_{1+}^{n}\right)e^{iw_{1,n}z}\mathbf{\digamma}\left[\intop_{0}^{z}e^{-iw_{1,n}z'}\mathbf{P}(\mathbf{r_{||}},z')dz'\right]_{\kappa_{n}}\nonumber \\
 & +\frac{ik_{0}^{2}}{2\varepsilon_{0}w_{1,n}}\left(\hat{\boldsymbol{\varrho}}{}_{1-}^{n}\otimes\hat{\boldsymbol{\varrho}}{}_{1-}^{n} \right)e^{-iw_{1,n}z}\mathbf{\digamma}\left[\intop_{z}^{h(\mathbf{r_{||}})}e^{iw_{1,n}z'}\mathbf{P}(\mathbf{r_{||}},z')dz'\right]_{\kappa_{n}}\nonumber \\
 & +\frac{ik_{0}^{2}}{2\varepsilon_{0}w_{1,n}}\left(r_{\textnormal{f}}^{\varrho, n} \hat{\boldsymbol{\varrho }}_{1+}^{n}\otimes\hat{\boldsymbol{\varrho}}{}_{1-}^{n} \right)e^{iw_{1,n}z}\mathbf{\digamma}\left[\intop_{0}^{h(\mathbf{r_{||}}}e^{iw_{1,n}z'}\mathbf{P}(\mathbf{r_{||}},z')dz'\right]_{\kappa_{n}}.\label{eq:Main}
\end{align}

The physical interpretation of the different terms in Eq.~(\ref{eq:Main}) can be found in literature\cite{Abass2017,Rottenberg:2011}. $\boldsymbol{\kappa}_{n}$ is the momentum component parallel to the interface. The index $n$ runs through all considered plane wave components. $\varepsilon_{0}$ is the vacuum permittivity. $h(\mathbf{r_{||}})$ is the scattering structure's height profile. $\hat{\boldsymbol{\varrho}}^n_{1+}$ and $\hat{\boldsymbol{\varrho}}^n_{1-}$ are unit vectors of upward and downward going waves in medium I. $k_{0}=2\pi/\lambda$ is the vacuum wavenumber, $w_{1,n}=\sqrt{\varepsilon_{1}k_{0}^{2}-\kappa_{n}^{2}}$
is the wave-vector component along the $z$ direction in medium I,
$r_{f}$ is the reflection coefficient for a planar system in the absence of a scattering structure in region II. The operator "$\mathbf{\digamma}$" denotes a Fourier transform of the respective quantity.  The incident field contributions are expressed by $ \mathbf{E}^{-,\mathrm{\varrho}}_{\mathrm{II,inc}}(\kappa_{n})=E_{\mathrm{inc}}^{\varrho}\hat{\boldsymbol{\varrho}}_{1-}^{n} $ and $ \mathbf{E}^{+,\mathrm{\varrho}}_{\mathrm{II,inc}}(\kappa_{n})= r_{\textnormal{f}}^{\varrho,n} E_{\mathrm{inc}}^{\varrho}\hat{\boldsymbol{\varrho}}_{1+}^{n} $ where $E_{\mathrm{inc}}^{\varrho}$ is the amplitude of the incident field. Equation (\ref{eq:Main}) is an implicit integral equation as the excess polarization $\mathbf{P}(\mathbf{r_{||}},z)$ is given by the product between the electric field in region II and the difference in permittivity relative to the background media I, which is expressed by
\begin{equation}
\mathbf{P}(\mathbf{r_{||}},z)=\begin{cases}
\varepsilon_{0}\Delta\varepsilon\mathbf{E}_{\mathrm{II}}(\mathbf{r_{||}},z) & \text{for }0\leq z\leq h(\mathbf{r_{||}}),\\
0 & \text{for }z<0\text{ and }z>h(\mathbf{r_{||}}).
\end{cases}\label{eq:Polarization}
\end{equation}
We note that as long as an exact $\mathbf{P}(\mathbf{r_{||}},z)$ is considered, Eq.~(\ref{eq:Main}) is also exact. 

In our previous publication\cite{Abass2017} we considered the surface texture to be only a weak perturbation to the flat interface. Therefore, we solve Eq.~(\ref{eq:Main}) by assuming an ansatz field in region II consisting of upward and downward propagating plane waves according to
\begin{align}
\mathbf{\mathbf{E}}_{\mathrm{II}}^{\varrho}(\mathbf{r_{||}},z)= & \sum_{n}e^{i\boldsymbol{\kappa}_n \cdot\mathbf{r_{||}}} \left(\mathbf{E}^{+,\varrho}_{\mathrm{II}}(\boldsymbol{\kappa}_n)e^{iw_{1,n}z}+\mathbf{E}^{-,\varrho}_{\mathrm{II}}(\boldsymbol{\kappa}_n)e^{-iw_{1,n}z}\right).\label{eq:Flatmodes-1}
\end{align}
This ansatz assumes that the field in region II can simply be expressed by up and down going plane waves in region I with field amplitudes not changing along $z$. This approximation can still be valid over an important surface amplitude perturbation regime as long as a proper averaging approximation in handling the third and fourth terms of Eq.~(\ref{eq:Main}) is used. By injecting this ansatz into in Eq.~(\ref{eq:Main}), one can subsequently obtain a linear equation system that solves for the unknown amplitudes of the up and down going plane waves in region II, $\mathbf{E}^{+,\varrho}_{\mathrm{II}}(\kappa_{n})$  and $\mathbf{E}^{-,\varrho}_{\mathrm{II}}(\kappa_{n})$, respectively. The usage of Eq.~(\ref{eq:Flatmodes-1}) as ansatz for the field in region II constrains the solution within the shallow surface amplitude regime even though it allows a computationally efficient forward and inverse modelling forward and inverse modelling of quasi-periodic textures comprising of incommensurable periodic components. 

To go beyond the shallow surface amplitude regime one must utilize a different ansatz to describe the field in region II. We choose an ansatz by first considering an alternative reference structure from which the final disordered system is considered as its perturbation. Instead of considering the disordered texture to be a perturbation relative to a flat interface, we consider the texture to be a perturbation relative to a certain ordered pattern. In such case, we can employ Bloch modes to describe the field in region II. Moreover, if we limit ourselves to binary type textures that maintain their shape along the vertical direction (as shown in Fig.~\ref{fig1}), then it is further justified to consider a single set of Bloch modes to represent the field in region II.  Thus, instead of the plane-wave ansatz in Eq.~(\ref{eq:Flatmodes-1}), we consider the field in the modulation region to be described by Bloch modes of the form
\begin{equation}
    \mathbf{\mathbf{E}}_{\mathrm{II}}^{\varrho}(\boldsymbol{r}_{||},z)=   \sum_{l}\left(  A^+_{l}  e^{iw_{l}z}  +  A^-_{l} e^{-iw_{l}(z-h_0)}\right) \sum_{n} \Tilde{\mathbf{E}}^{\varrho}_{l,\mathrm{II}}(\boldsymbol{\kappa}_{n}) e^{i\boldsymbol{\kappa}_{n}\cdot\boldsymbol{r}_{||} }, \label{eq:Bloch}
\end{equation}
where $A^\pm_{l}$ is the amplitude of the up and down going component of mode $l$, $w^l$ is the mode propagation constant along the height of the texture, $\mathbf{E}^{\varrho}_{l,\mathrm{II}}(\boldsymbol{\kappa}_{n})$ are the Fourier components of Bloch mode $l$ with polarization $\varrho$ , and $h_0$ is the height of the binary texture\cite{Li:96}. Both $w^l$ and $\mathbf{E}^{\varrho}_{l,\mathrm{II}}(\boldsymbol{\kappa}_{n})$ are obtained by performing eigenmode calculations for the periodic unperturbed reference structure, similar to what is done in the initial step of RCWA\cite{Li:97,Popov:01,Caballero2012}. As these Bloch modes are well defined for periodic systems, a full-wave solution can be recovered independently of the binary texture height as long as the proper set of eigenmodes is considered. The Green's tensor scattering formalism however, allows one to utilize non-native Bloch modes as ansatz at the expense of accuracy. One can essentially consider different periodic reference structures from which the Bloch modes are taken, to flexibly manage the trade-off between accuracy and computational costs in addressing extended domains. For example, as sketched in Fig.~\ref{fig1}, one can use a simpler unperturbed periodic reference structure (blue cylinder arrays) to obtain an approximate scattering response (blue wavefront) for the actual disordered system (purple wavefront). Utilizing modes of a simpler periodic system translates into considering a smaller number of plane wave components to represent the field in region II, which in turn leads to a smaller matrix to invert. However, as will be shown in the next section, this does not necessarily mean that one loses information of scattering to plane wave components with tangential wave-vectors not considered in deducing the field in region II.

By inserting Eq.~(\ref{eq:Bloch}) into Eq.~(\ref{eq:Main}) and evaluating the integral terms at different positions along $z$, one can obtain a complete linear equation system to deduce $A^\pm_{l}$. In principle, one can choose any two $z$ positions in region II to obtain enough equations for all $2l$ unknowns. For convenience, we assume $z=0$ and $z=h_0$ since this leads to simpler terms. Evaluating Eq.~(\ref{eq:Main}) at $z=h_0$ one obtains 
\begin{align}
\sum_{l}\left(  A^+_{l}  e^{iw_{l}h_0}  +  A^-_{l} \right) \tilde{\mathbf{E}}^{\varrho}_{l}(\boldsymbol{\kappa}_n) = & E_{\mathrm{inc}}^{\varrho}(\boldsymbol{\kappa}_{n})\left( \hat{\boldsymbol{\varrho}}_{1-}^{n} e^{-iw_{1,n}h_0} + r_{\textnormal{f}}^{\varrho,n} \hat{\boldsymbol{\varrho}}_{1+}^{n} e^{iw_{1,n}h_0} \right) \nonumber \\ 
& +  \Delta\varepsilon  \sum_{\pm'} \sum_{l} A^{\pm'}_{l} \sum_{m} \tilde{\boldsymbol{\Upsilon}}^{n,\varrho,\mathrm{top}}_{l,\pm',m} \cdot\: \sum_{\varrho'}   \tilde{\mathbf{E}}^{\varrho'}_{l}(\boldsymbol{\kappa}_n-\mathbf{G}_{m}) ,\label{eq:FieldII_zh0}
\end{align}
where $\tilde{\boldsymbol{\Upsilon}}^{ n,\varrho,\mathrm{top}}_{l,\pm',m} $ is a dyadic coupling coefficient between plane wave components expressed by
\begin{align}
\tilde{ \boldsymbol{\Upsilon}}^{n,\varrho,\mathrm{top}}_{l,\pm',m} = &  \frac{ik_{0}^2}{2w_{1,n}} e^{iw_{1,n}h_0}  e^{iw_{l}(h_0\mp h_0)/2} \mathbf{U}^\varrho\otimes\left(\hat{\boldsymbol{\varrho}}_{1+}^{n} \digamma\left[  \frac{ e^{iw^{n,+}_{l,\pm'} h( \mathbf{r_{||} } ) }-1}{iw^{n,+}_{l,\pm'}}\right] +r_{\textnormal{f}}^{\varrho,n} \hat{\boldsymbol{\varrho}}_{1-}^{n} \digamma\left[  \frac{ e^{iw^{n,-}_{l,\pm'} h( \mathbf{r_{||} } ) }-1}{iw^{n,-}_{l,\pm'}}\right] \right)_{\mathbf{G}_m}, 
 \label{eq:Upsilon1}
\end{align}
with $\tilde{w}_{l,\pm}^{n,+}= w^{l}_{n} \mp w_{1,n} $. If one decomposes the Bloch modes in terms of plane waves with $p$ and $s$ polarization cases, one has $\mathbf{U}^\mathrm{p} = \left( \frac{1}{\widetilde{\varepsilon}} \frac{\kappa_{n}}{k_{0}n_{1}}  \hat{\boldsymbol{z}} \mp \frac{w_{1,n}}{k_{0}n_{1}} \hat{\boldsymbol{\kappa}}^n \right)$ and $\mathbf{U}^\mathrm{s} = \mathbf{\hat{s}}^n $ respectively.  

At this point, we would like to first discuss in more detail the physics behind Eq.~(\ref{eq:FieldII_zh0}) and Eq.~(\ref{eq:Upsilon1}). The first two terms of Eq.~(\ref{eq:FieldII_zh0}) are the driving contribution of the incident field in the absence of a scattering structure in region II. Note that as the up and down going portions of the driving contribution are not eigenmodes of the system in region II, they each contribute to the generation of up and down going waves. 

The second line of Eq.~(\ref{eq:FieldII_zh0}) describes the coupling between different plane wave components which comprise the available modes in region II, much like what one would encounter in RCWA. Though involving many factors, one can clearly distinguish them and their impact on the coupling interactions. We write $\Delta \varepsilon$ apart from $\tilde{\boldsymbol{\Upsilon}}^{ n,\varrho,\mathrm{top}}_{l,\pm} $ to clearly show that the permittivity contrast is proportional to the coupling strength between plane wave components. The remaining coupling coefficient $\tilde{\boldsymbol{\Upsilon}}^{ n,\varrho,\mathrm{top}}_{l,\pm} $ encompasses the impact of modal and geometrical properties on the scattering system. 

The modal properties that influence $\tilde{\boldsymbol{\Upsilon}}^{ n,\varrho,\mathrm{top}}_{l,\pm} $ are the eigenfield profiles and the propagation constants of the modes involved in the scattering process ($w_l$). The dependence on the eigenfield profile distributions in particular is encompassed by the field polarization terms $\varrho$ of the plane wave components involved ($\mathbf{U}^\mathrm{p}$, $\hat{\boldsymbol{\varrho}}$). The dependence on the scattering geometry is contained by the terms involving Fourier transforms of exponentials $h( \mathbf{r_{||} } )$. The background system in which our scattering texture resides in impacts $\tilde{\boldsymbol{\Upsilon}}^{ n,\varrho,\mathrm{top}}_{l,\pm} $ through the reflection coefficient $r_{\textnormal{f}}$ and the propagation constant ($w_{1,n}$). Depending on the phase of $r_{\textnormal{f}}$, one can essentially enhance the coupling between plane waves by placing the scattering structure on top of a strongly reflecting surface as implied by Eq.~(\ref{eq:Upsilon1}). In principle one can consider a multilayer stack instead of a substrate as long as $r_{\textnormal{f}}$ corresponds to that multilayer stack, which is done in this work. 

Having discussed the physics, we proceed to obtain the next set of equations by evaluating Eq.~(\ref{eq:Main}) at $z=0$, which would lead to 
\begin{align}
\sum_{l}\left(  A^+_{l}   +  A^-_{l} e^{-iw_{l}h_0}  \right) \tilde{\mathbf{E}}^{\varrho}_{l}(\boldsymbol{\kappa}_{n}) = & E_{\mathrm{inc}}^{\varrho} (\boldsymbol{\kappa}_{n}) \left( \hat{\boldsymbol{\varrho}}_{1-}^{n} + r_{\textnormal{f}}^{\varrho,n} \hat{\boldsymbol{\varrho}}_{1+}^{n} \right)   \nonumber \\ 
& + \Delta\varepsilon   \sum_{\pm'} \sum_{l} A^{\pm'}_{l} \sum_{m}  \tilde{ \boldsymbol{\Upsilon}}^{n,\varrho,\mathrm{bottom}}_{l,\pm',m} \cdot\: \sum_{\varrho'}  \tilde{\mathbf{E}}^{\varrho'}_{l}(\boldsymbol{\kappa}_{n}-G_{m}) ,\nonumber\\\label{eq:FieldII_z0}
\end{align}
where $ \tilde{ \boldsymbol{\Upsilon}}^{n,\varrho,\mathrm{bottom}}_{l,\pm',m} $ is expressed by 
\begin{align}
\tilde{ \boldsymbol{\Upsilon}}^{n,\varrho,\mathrm{bottom}}_{l,\pm',m} = &  \frac{ik_{0}^2}{2w_{1,n}} e^{iw_{l}(h_0 \mp h_0)/2} \mathbf{U}^\varrho\otimes \left( \hat{\boldsymbol{\varrho}}_{1-}^{n} \digamma\left[  \frac{ e^{iw^{n,-}_{l,\pm'} h( \mathbf{r_{||} } ) }-1}{iw^{n,-}_{l,\pm'}}\right] + r_{\textnormal{f}}^{\varrho,n} \hat{\boldsymbol{\varrho}}_{1-}^{n} \digamma\left[  \frac{ e^{iw^{n,-}_{l,\pm'} h( \mathbf{r_{||} } ) }-1}{iw^{n,-,}_{l,\pm'}}\right] \right)_{\mathbf{G}_m} 
 \label{eq:Upsilon2}
\end{align}

Together Eqs.~(\ref{eq:FieldII_zh0}) and (\ref{eq:FieldII_z0}) form a complete linear equation system from which $A^\pm_{l}$ can be solved for with matrix inversion. We stress again that as long as the binary structures are unchanged along $z$ and the proper set of modes are considered,  Eqs.~(\ref{eq:FieldII_zh0}) and (\ref{eq:FieldII_z0}) will provide a rigorous analytical solution to the problem. 

\section{Scattering of disordered structures}

As mentioned before, the field in other regions can be deduced once the field in region II is known. One way to do this is by exploiting the continuity of the tangential components at the interface between regions as would be done in RCWA. By doing so, one consider in the other regions the same amount of plane waves used to describe the modes in the scattering region (Region II). This is of course not a concern if one considered the actual native eigenmodes of the scattering structure. For the purpose of approximate calculations of extended disordered systems however, we wish to utilize as ansatz Bloch modes that are not native to the scattering structure to reduce computational costs. Non-native Bloch modes may not naturally couple to all radiation channels to which the actual scattering structure has access. To approximately deduce the scattering strength of all channels accessible by the considered structure therefore, one must utilize another approach. We accomplish this by again utilizing a Green's tensor scattering formalism similar to Eq.~(\ref{eq:Main}). For example, the transmitted field $\mathbf{E}^{\varrho}_{\mathrm{t}}$ can be written in the form 
\begin{equation}
\mathbf{E}^{\varrho}_{\mathrm{t}}(\boldsymbol{\kappa}_{n'}) = t_{\textnormal{f}}^{\varrho,n} E_{\mathrm{inc}}^{\varrho}(\boldsymbol{\kappa}_{n})\hat{\boldsymbol{\varrho}}_{3-}^{n'} +\frac{ik_{0}^{2}}{2\varepsilon_{0}w_{1,n'}}\left(t_{\textnormal{f}}^{\varrho, n} \hat{\boldsymbol{\varrho }}_{3-}^{n'}\otimes\hat{\boldsymbol{\varrho}}{}_{1-}^{n'} \right) \mathbf{\digamma}\left[\intop_{0}^{h(\mathbf{r_{||}})}e^{iw_{1,n'}z'}\mathbf{P}(\mathbf{r_{||}},z')dz'\right]_{\boldsymbol{\kappa}_{n'}}.\label{eq:RegIII}
\end{equation}
Injecting Eq.~(\ref{eq:Bloch}) and known values of $A^\pm_{l}$ into  Eq.~(\ref{eq:RegIII}), one would obtain
\begin{align}
\mathbf{E}^{\varrho}_{\mathrm{t}}(\boldsymbol{\kappa}_{n'})=& t_{\textnormal{f}}^{\varrho,n'} E_{\mathrm{inc}}^{\varrho} (\boldsymbol{\kappa}_{n}) \hat{\boldsymbol{\varrho}}_{3-}^{n'} + \Delta\varepsilon   \sum_{\pm'} \sum_{l} A^{\pm'}_{l} \sum_{m} \tilde{ \boldsymbol{\Theta}}^{n',\varrho}_{l,\pm',m} \cdot\:  \sum_{\varrho'}  \tilde{\mathbf{E}}^{\varrho'}_{l}(\boldsymbol{\kappa}_{n'}-G_{m}),\label{eq:ReqIIIlin}
\end{align}
where $\tilde{ \boldsymbol{\Theta}}^{n',\varrho,\mathrm{t}}_{l,\pm',m}$ is 
\begin{align}
\tilde{ \boldsymbol{\Theta}}^{n',\varrho,\mathrm{t}}_{l,\pm',m} = &  \frac{ik_{0}^2}{2w_{1,n'}}  e^{iw_{l}(h_0 \mp h_0)/2} \hat{\boldsymbol{\varrho}}_{3-}^{n'}\otimes \left( t_{\textnormal{f}}^{\varrho,n} \hat{\boldsymbol{\varrho}}_{1-}^{n'} \digamma\left[  \frac{ e^{iw^{n',-}_{l,\pm'} h( \mathbf{r_{||} } ) }-1}{iw^{n',-,}_{l,\pm'}}\right] \right)_{\mathbf{G}_m}. \label{eq:Theta}
\end{align}
We utilize $n'$ instead of $n$ in Eqs.~(\ref{eq:ReqIIIlin}) and (\ref{eq:Theta}) to highlight the fact that the plane wave components considered outside of region II does not need to be the same as the ones considered in region I. With our Green's scattering formalism, the scattered amplitude of channel $n'$ can still be calculated even though the Bloch modes used as ansatz for the field in region II does not contain plane wave components with the particular $\boldsymbol{\kappa}_{n'}$ so long there the scattering texture can supply a momentum $\mathbf{G}_m$ such that  $\boldsymbol{\kappa}_{n'}=\boldsymbol{\kappa}_n-\mathbf{G}_m$

Similarly, the reflected field above the surface texture would be of the form
\begin{align}
\mathbf{E}^{\varrho}_{\mathrm{r}}(\kappa_{n})=& r_{\textnormal{f}}^{\varrho,n} E_{\mathrm{inc}}^{\varrho} (\boldsymbol{\kappa}_{n}) \hat{\boldsymbol{\varrho}}_{1+}^{n} + \Delta\varepsilon   \sum_{\pm'} \sum_{l} A^{\pm'}_{l} \sum_{m} \tilde{ \boldsymbol{\Theta}}^{n,\varrho,\mathrm{r}}_{l,\pm',m} \cdot\:  \sum_{\varrho'}  \tilde{\mathbf{E}}^{\varrho'}_{l}(\boldsymbol{\kappa}_{n}-\mathbf{G}_{m}),\label{eq:ReqIlin}
\end{align}
where $\tilde{ \boldsymbol{\Theta}}^{n,\varrho,\mathrm{r}}_{l,\pm'}$ is 
\begin{align}
\tilde{ \boldsymbol{\Theta}}^{n,\varrho,\mathrm{r}}_{l,\pm',m} = &  \frac{ik_{0}^2}{2w_{1,n}}  e^{iw_{l}(h_0 \mp h_0)/2} \hat{\boldsymbol{\varrho}}_{1+}^{n} \otimes \left( \hat{\boldsymbol{\varrho}}_{1-}^{n} \digamma\left[  \frac{ e^{iw^{n,-}_{l,\pm'} h( \mathbf{r_{||} } ) }-1}{iw^{n,-}_{l,\pm'}}\right] + r_{\textnormal{f}}^{\varrho,n} \hat{\boldsymbol{\varrho}}_{1-}^{n} \digamma\left[  \frac{ e^{iw^{n,-}_{l,\pm'} h( \mathbf{r_{||} } ) }-1}{iw^{n,-,}_{l,\pm'}}\right] \right)_{\boldsymbol{\kappa}_m-\boldsymbol{\kappa}_n}. \label{eq:Theta2}
\end{align}
\end{widetext}

One disadvantage of this method, however, is that energy conservation can be violated during our scattering calculations. This is fundamentally due to our choice for the ansatz form, which is not the eigensolution of the actual scattering geometry. A larger violation of energy conservation will occur the further the considered scattering structure is from the reference structure from which the assumed ansatz is taken from. That is , however, not to say that one cannot obtain a decent estimate of the scattering properties as it is. Upon forcing energy conservation through renormalization of the obtained diffracted powers, however, one further extends the regime where the method can provide a good estimate for the scattering response of perturbed systems. Here, the renormalization is done by simply dividing the deduced power quantities (transmittance, reflectance, scattered power of particular diffraction orders, and absorption) by the sum of all possible diffraction orders and absorption in the system.

The construction of the matrix system described by Eqs.~(\ref{eq:FieldII_zh0}) and (\ref{eq:FieldII_z0}) can be paralelized to reduce computational time and memory usage. We note that Fourier transform calculations involved in terms with  $ \digamma\left[  \frac{ e^{iw h( \mathbf{r_{||} } ) }-1}{iw}\right]_{\mathbf{G}_m} $ may be expensive. This is especially so if one performs the calculation for every possible coupling and scattering interaction. Deducing the Fourier transform of such exponential functions, however, can be greatly simplified in different ways. One possible approach is considering a taylor expansion of these exponents, which allows one to calculate only the Fourier transform of the polynomial functions ($ h( \mathbf{r_{||} } )$, $ h( \mathbf{r_{||} } )^2$, and so on) that can be used to reconstruct the actual coupling coefficient strength. Another possible approach, which is used here, is by taking advantage on the fact that we are considering only binary structures. This allows one to instead consider the Fourier transform of $ \frac{( e^{iw h_0 }-1)}{iw}\digamma\left[ h( \mathbf{r_{||} } )/h_0 \right]_{\mathbf{G}_m} $.   
   
\section{Benchmarking}

In order to describe how well our approach can address disordered binary surface textures, we here discuss the scattering of light by 4 by 4 2D square prism dielectric grating supercells with varying randomness of unit cell fill factor $FF=l/P$, where $P$ is the single unit cell period and $l$ is the sidelength of the square prisms. We compare our method with RCWA (used here as benchmark), which provides a rigorous solution for the binary system as long as enough plane waves are used to describe the Bloch modes.
 
\begin{figure}[ht]

	\includegraphics[scale=0.3]{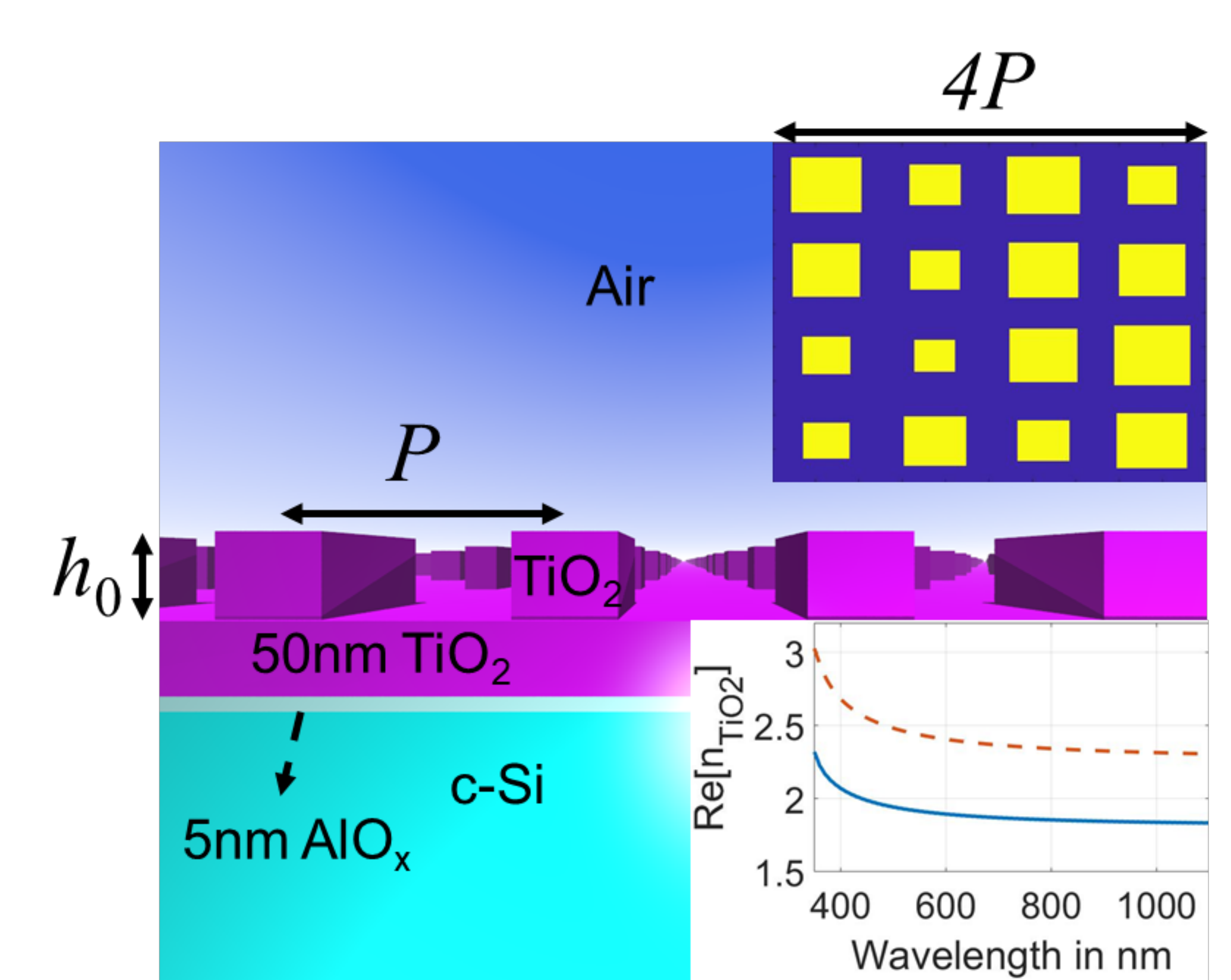}
	\caption{\textit{ A system of 4 by 4 TiO$_2$ grating supercell on top of a 20~nm TiO$_2$/5~nm AlO$_\mathrm{x}$/ c-Si substrate multilayer stack with individual period  $P=500$~nm. The top right inset figure gives a 4 by 4 example with $\Delta FF=0.33$. The bottom right inset gives the two refractive index of TiO$_2$ considered in the benchmarking. The higher (dashed line) and lower (solid line) refractive index material pertains to TiO$_2$ deposited with high and low temperature, respectively. }}\label{fig2}

\end{figure}
Each of cell is considered to have individual period of $P=500$~nm, thus leading to a total supercell period size of 2~$\mu$m. To be more specific, we consider the case of TiO$_2$ binary grating supercells placed on top of a air/TiO$_2$/AlO$_\mathrm{x}$/c-Si substrate layer stack as shown in Fig.~\ref{fig2} at normal incidence. Such layer stack is relevant for interdigitated back contact cell configurations, where there are only passivating layers on the front side where sun light enters\cite{Spinelli2013}. The corresponding reflection and transmission coefficient of the layer stack are deduced via Transfer Matrix Method calculations\cite{Born1980}. For the benchmarking calculations, we consider both high and low index TiO$_2$ (bottom right inset of Fig.~\ref{fig2}) pertaining to high and low temperature deposition respectively. We further consider two grating height cases of 100 and 150~nm. The material and geometry combination of the grating structure are of physical relevance for electrically decoupled light trapping structure schemes to enhance light trapping and incoupling in solar cells\cite{Isabella:16}.  

We inspect the scattering response both into air and into the c-Si substrate in the wavelength range of 400-1100~nm, which is relevant for c-Si solar cells. Note that the height and index combination cause the grating system to go beyond the shallow perturbation regime for the considered wavelength range.
 
The top right inset of Fig.~\ref{fig2} provides a top view of one realization of the 4 by 4 supercell system. These 4 by 4 supercells were generated by employing a flat probability distribution within varied ranges around the center fill factor value $FF_\mathrm{c}=0.5$. We proceed below to compare the scattering characteristics of supercells with varying unit cell fill factor distribution range $\Delta FF=FF_{max}-FF_{min}$, obtained both with GMoUB and with RCWA. We wish to stress however, that one can also consider other types of disorders (eg. shape, positions, materials).

For all our GMoUB calculations given below, we employ the Bloch modes ansatz of the unperturbed grating structure with uniform unit cell fill factors. The reference unit cell fill-factor is taken to be the mean of every considered instance, which varies only slightly from 0.5.  Each Bloch mode comprises of 25 by 25 plane waves (12 plane waves in $\pm$ $x$ and $y$ directions including the zeroth order) with parallel momentum $k_{||,m,n} = m G \hat{x}+ n G \hat{y}$ where $G=2\pi/P$. The RCWA reference calculations are done considering all possible plane wave components that can couple to radiation in either region I and III (Fig.~\ref{fig1}. A sketch of the k-space mesh indicating the plane wave components taken into account by GMoUB and RCWA in deducing the field in region II is shown in the appendix (Fig.~\ref{fig8}). The GMoUB calculations essentially utilize a more sparse k-space mesh (blue filled circles) as compared to the RCWA calculations (red open circles) in deducing the field in region II.

\begin{figure}[ht]

	\includegraphics[scale=0.45]{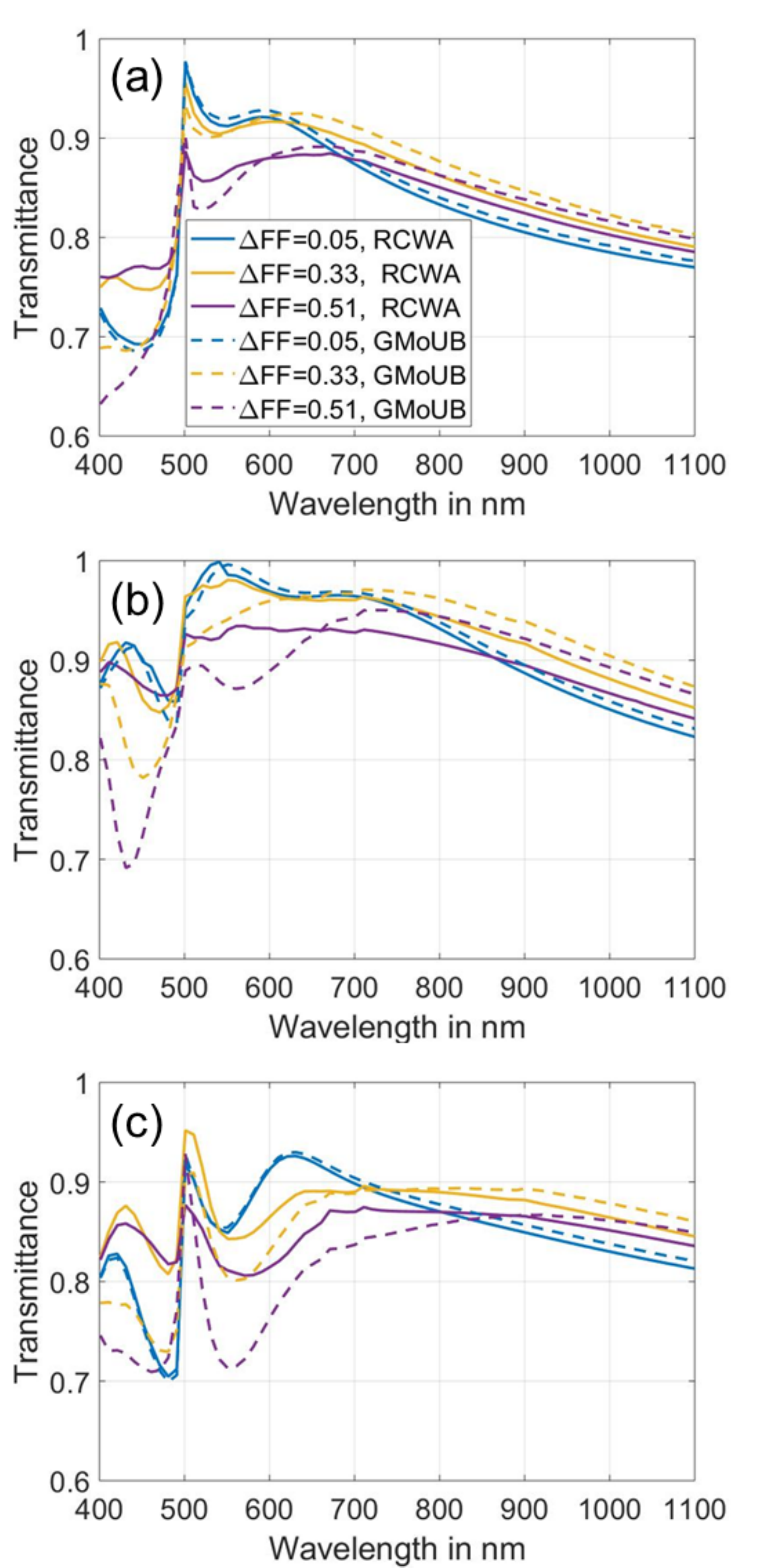}
	\caption{\textit{Comparison of transmittance spectra for different supercells obtained with RCWA and GMoUB for the case of $h_0=100$~nm (a) low index and (b) high index TiO$_2$ and also (c) $h_0=150$~nm for low index TiO$_2$. }}\label{fig3}

\end{figure}
Figure~\ref{fig3}  provides a comparison of total transmittance spectra into the c-Si substrate calculated with our GMoUB formalism and with RCWA for three fill factor standard deviation values and different grating height and material configurations. For all grating height and material configuration cases, the trends in the transmittance spectra as fill factor disorder is increased are well captured by our GMoUB method. For example, if one inspects Fig.~\ref{fig3}a for $h_0=100$~nm and low index TiO$_2$, one can see that both methods show an increase of transmittance at longer wavelengths and decrease at shorter waveelngths. The discrepancy between our method and RCWA is smaller at longer wavelengths. Shorter wavelengths perceive the disordered perturbations more strongly as Bloch modes in this regime perceive a larger optical path length in the scattering region (region II). Therefore, error in the k-vector due to our usage of a non-native ansatz can translate into a larger phase error. 

When one increases the refractive index (Fig.~\ref{fig3}b), one can see that there is less agreement with increasing disorder between our approximate GMoUB method and RCWA, except again at longer wavelengths. Whereas the low index case sustains excellent agreement between both methods even for $\Delta FF=0.51$ down to a wavelength of 700~nm, there is visibly less agreement in the  400-700~nm wavelength range upon increase of refractive index. This is due to the higher refractive index of TiO$_2$ which increases the propagation constant of the Bloch modes, rendering the calculation of perturbed systems prone to larger optical path length errors when a non-native eigenmode is utilized as the ansatz. A similar picture occurs when one increases the height of the grating region (Fig.~\ref{fig3}c). Such a system also increases the possibility of larger optical path length error, which then would shift the regime where the method provides excellent agreement to even longer wavelengths. 

\begin{figure}[ht]

	\includegraphics[scale=0.2]{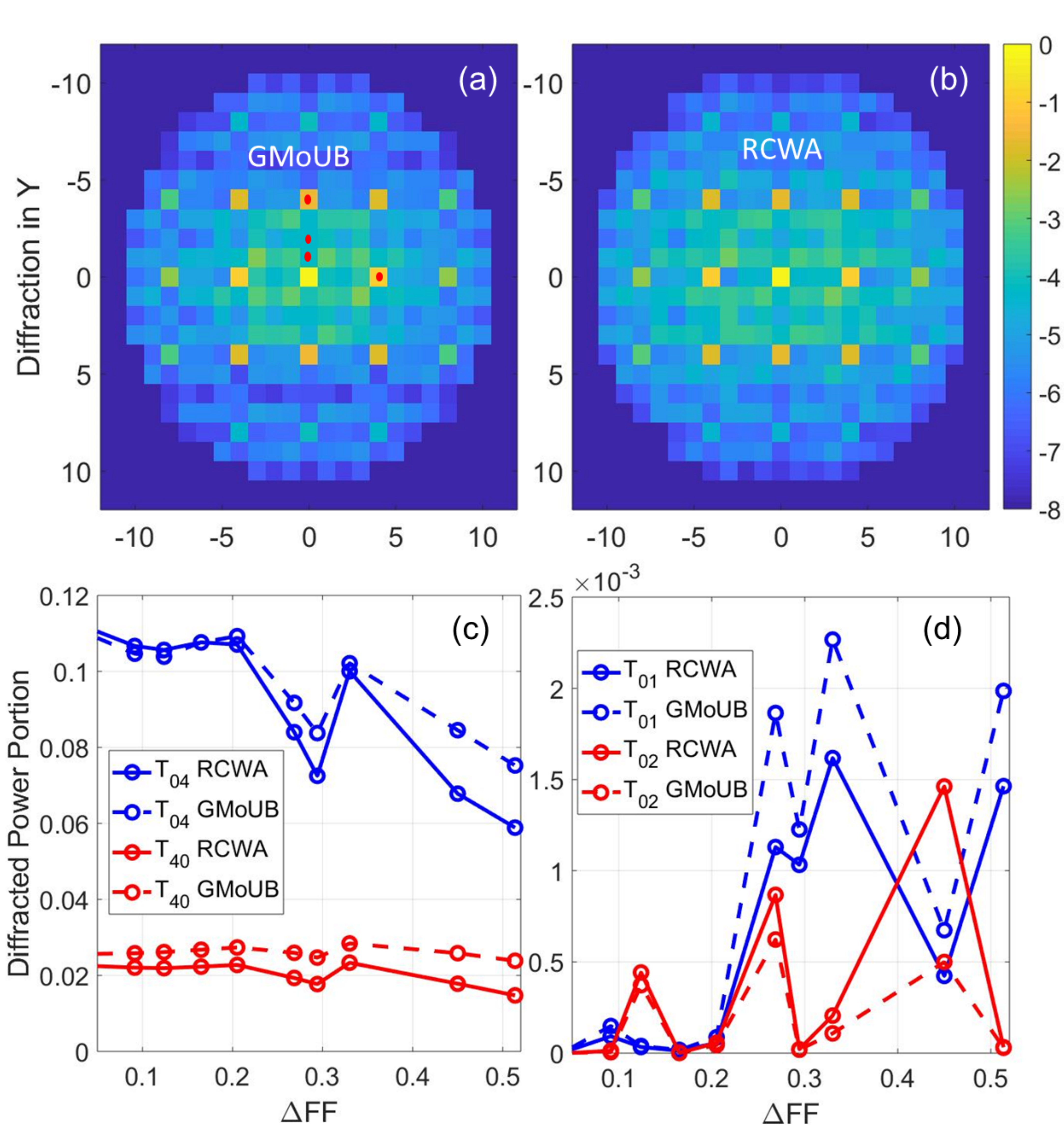}
	\caption{\textit{Transmitted diffracted power  distribution into c-Si substrate in logarithmic scale ($\mathrm{log}_{10}$) in k-space at wavelength of 700~nm for the system with high index TiO$_2$, $h_0=100$~nm, and $\Delta FF = 0.33$ obtained with (a) GMoUB and (b) RCWA. Power sent to the diffraction orders pertaining to (c) a dominant diffraction order that already occurs in the unperturbed system and (d) that only arises in the supercell system due to the fill factor perturbation as a function of $\Delta FF$. The red dots in (a) indicate the diffraction orders that are plotted in (c) and (d). }}\label{fig4}

\end{figure}

In Fig.~\ref{fig4}(a), we show the k-space resolved distribution of the diffraction scattering in the transmission direction in the c-Si substrate at $\lambda= 700$nm. A good agreement between our GMoUB method (a) and RCWA (b), can be seen. Diffraction contributions from original orders existing already in the unperturbed system and newly arising contributions due to the introduced fill-factor disorder are all well estimated. To show this even more clearly, we plot the evolution of the diffracted power as a function of  $\Delta FF$. Figure \ref{fig4}(c) gives a plot of two transmission diffraction order $T_{04}$ and $T_{40}$ diffraction orders in dependence on $S_{\mathrm{D}}$ at $\lambda= 700$nm. These diffraction orders already exist in the unperturbed system and hence their relatively large values. They essentially correspond to the first order diffraction processes in the y and x directions respectively (indicated by the red circle dots in Fig.~\ref{fig4}(a)). Figure \ref{fig4}(d), in contrast, provides a similar plot but for $T_{01}$ and $T_{02}$ (corresponding to red square dots in Fig.~\ref{fig4}(a). These diffraction orders arise due to the introduced disorder. As can be seen in Fig.~\ref{fig4} (c) and (d), the evolution of the distributed power to each of the diffraction orders as $\Delta FF$ is well captured even up to fairly high $\Delta FF = 0.51$.

\begin{figure}[ht]

	\includegraphics[scale=0.31]{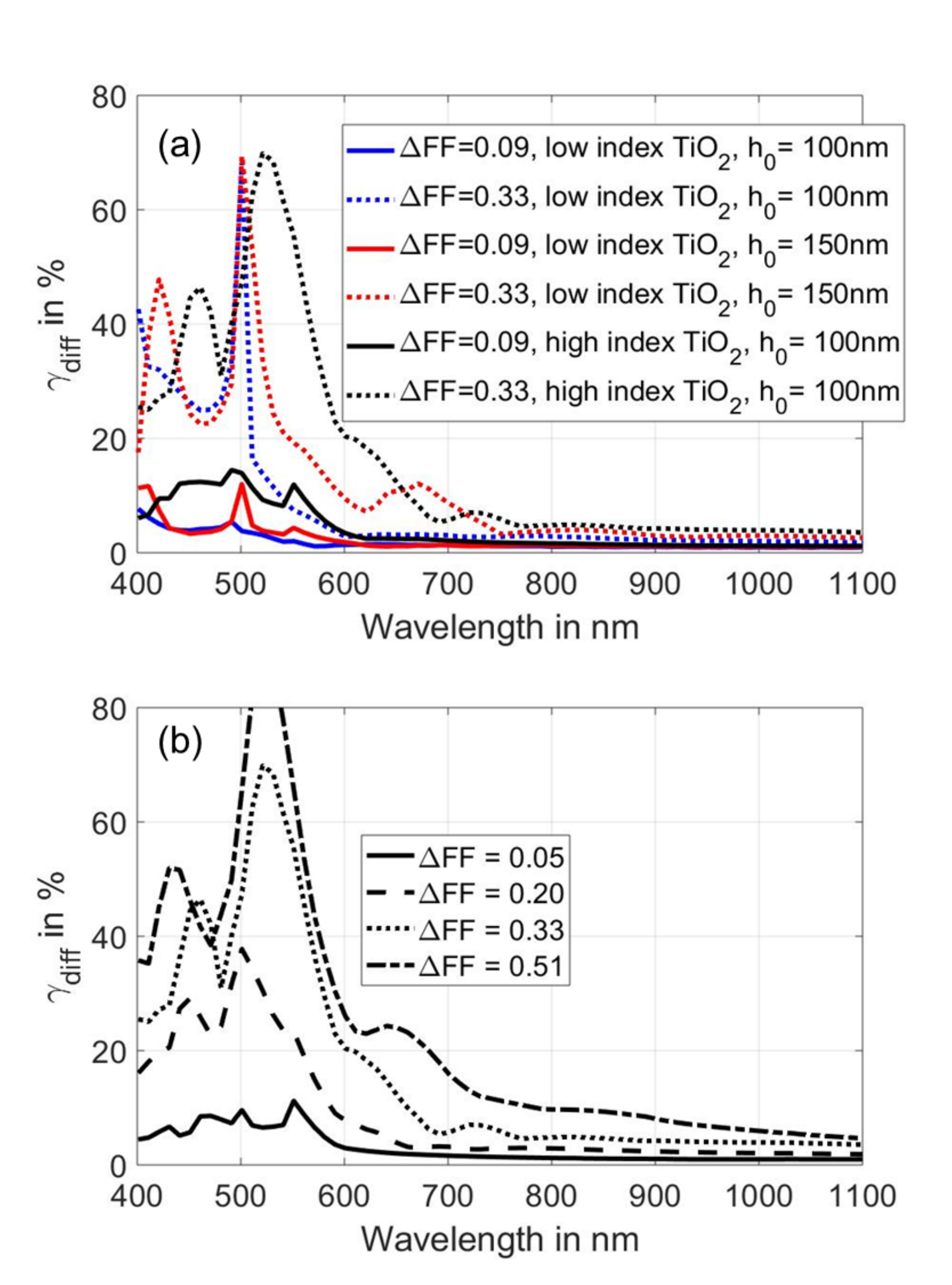}
	\caption{\textit{Plots of diffraction error figure-of-merit $\gamma_\mathrm{diff}$ as a function of wavelength for (a) different grating height and TiO$_2$ indeces for $\Delta FF=0.05$ and $\Delta FF=0.33$ and (b) for the case of of high index TiO$_2$ with $h_0 = 100$~nm for a more resolved change of $\Delta FF$.   }}\label{fig5}

\end{figure}
In order to describe the accuracy of our formalism, we define a figure-of-merit $\gamma_\mathrm{diff}$, which essentially gives the relative power scattered error to the various directions weighted to the portion of power scattered in each direction. $\gamma_\mathrm{diff}$ is mathematically described by
\begin{equation}
\gamma_\mathrm{diff}=\sum_{i,j}\frac{T^{\mathrm{ref}}_\mathrm{i,j}}{T^{\mathrm{ref}}_{\mathrm{tot}}}\frac{|T_\mathrm{i,j} - T^{\mathrm{ref}}_\mathrm{i,j}|} {T^{\mathrm{ref}}_\mathrm{i,j}}, \label{eq:Errordetailed}
\end{equation}
where $T_{ij}$ is the power sent to the transmission diffraction order $i,j$. The superscript label $^{\mathrm{ref}}$ refers to quantities calculated with RCWA, which are used as benchmark. Plots of $\gamma_\mathrm{diff}$ as a function of wavelength for selected configurations of increasing perturbation are given in Fig.~\ref{fig5}. As was also seen in Fig.~\ref{fig3}, the error at shorter wavelengths is generally larger and increases for larger grating feature height and refractive index (Fig.~\ref{fig5}(a)). They are also more sensitive to the perturbation.  This is in line with what we discussed above as shorter wavelengths may experience a larger phase error. Though $\gamma_\mathrm{diff}$ can actually reach fairly large values above $40\%$ at shorter wavelengths, especially when the grating index contrast is increased, we note that the total transmittance can still be in quite good agreement as seen in Fig.~\ref{fig3}. This means that the discrepancy lies in the detailed picture of power distribution of the different diffraction orders.  As mentioned before, the grating structure considered here is beyond the shallow amplitude regime. In the appendix, we show $\gamma_\mathrm{diff}$ from calculations utilizing only plane waves as ansatz (Eq.~\ref{eq:Flatmodes-1}) to consider the case with minimal fill factor perturbation ($\Delta FF=0.05$) for $h_0 = 100$~nm with either low and high TiO$_2$ refractive index. The usage of a simple plane wave ansatz leads to significantly larger errors in the diffraction distribution(Fig.~\ref{fig9}) especially at shorter wavelengths where one goes further beyond the shallow amplitude regime. A small error could still be obtained with the plane wave ansatz only at longer wavelengths where one approaches the shallow amplitude regime again. With the usage of Bloch modes, one can maintain a low error throughout the whole considered wavelength range. 

Figure~\ref{fig5}(b), provides a more detailed evolution of the $\gamma_\mathrm{diff}$ as the fill-factor disorder is increased for the case of high index TiO$_2$ with $h_0 = 100$~nm. We chose to focus on this system here because it provides the best anti-reflection performance of all the systems we consider. We note that for fairly strong perturbation  $\Delta FF=0.51$, $\gamma_\mathrm{diff}$ is maintained relatively low for wavelengths $>700$~nm, which is still a very relevant region for c-Si solar cells. This indicates that our approximate GMoUB formalism can provide a good and quick estimate of the disorder impact on relevant systems for solar cells’ light trapping and incoupling.

\begin{figure}[h]

	\includegraphics[scale=0.17]{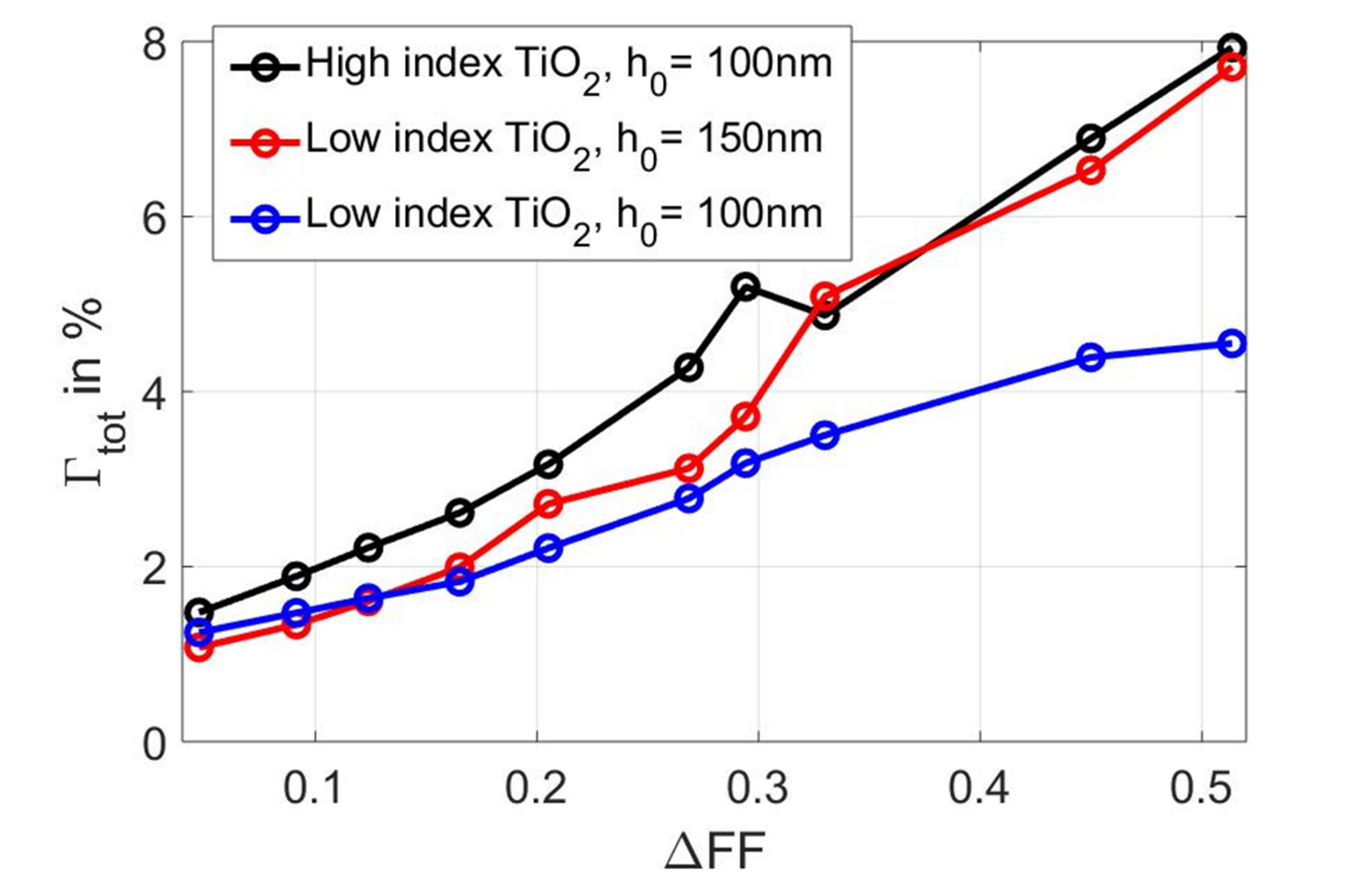}
	\caption{\textit{ Figure-of-merit for integrated reflectance and transmittance error $\Gamma_\mathrm{tot}$ as a function of $\Delta FF$ for different grating height and material configurations. }}\label{fig6}

\end{figure}

When one's concerns are macroscopic integrated quantities instead of an accurate picture of the angular distribution of scattered light, our formalism may provide an excellent prediction across a larger perturbation range than implied by $\gamma_\mathrm{diff}$. To demonstrate this, we consider another figure-of-merit $\Gamma_\mathrm{tot}$ which measures the total transmittance and reflectance error in the whole wavelength range of 400-1100~nm. $\Gamma_\mathrm{tot}$ is mathematically described by
\begin{equation}
\Gamma_\mathrm{tot}=\int \frac{ \sum T_\mathrm{i,j}(\lambda) -\sum T^{\mathrm{ref}}_\mathrm{i,j}(\lambda)+ \sum R_\mathrm{i,j}(\lambda) - \sum R^{\mathrm{ref}}_\mathrm{i,j}(\lambda)} {T^{\mathrm{ref}}_{\mathrm{int}}+R^{\mathrm{ref}}_{\mathrm{int}}} d\lambda .\label{eq:ErrorInt}
\end{equation}
where $T^{\mathrm{ref}}_{\mathrm{int}}$ and $T^{\mathrm{ref}}_{\mathrm{int}}$ are the total transmittance and reflectance integrated over the wavelength range of 400-1100~nm.
This figure-of-merit is essentially blind to error in the angular distribution of scattered power since it only concerns with the relative total transmittance and reflectance averaged over the whole wavelength range. Figure~\ref{fig6} provides a plot of $\Gamma_\mathrm{tot}$ as a function $\Delta FF$. Note the especially low error ($<8\%$) maintained for the entire perturbation regime for all different grating configurations adopted. Consistently, the grating system with the lowest refractive index and grating height exhibits the smallest $\Gamma_\mathrm{tot}$. We thus show that our perturbation formalism can provide a good estimate for macroscopic quantities such as total reflectance and transmittance. For applications such as light trapping in solar cells, macroscopic quantities such as total reflectance or expected short circuit current are the ones that actually matter instead of the accurate angular distribution picture. At small $\Delta FF$, we note that $\Gamma_\mathrm{tot}$ goes to a small number but not to 0. This lack of perfect convergence between GMoUB and our RCWA benchmark, is due to the limited amount of plane waves utilized to represent the Bloch modes. These two independent methods have a different convergence rate with respect to the number of plane waves utilized in the calculations.  
\section{Example large area disorder calculations for light trapping} 

Having benchmarked our method, we proceed to demonstrate its usage in large disordered domains. Instead of the 4 by 4 systems of the previous section, we consider here for example 20 by 20 supercell grating systems, which typically requires a large computational effort (either time, CPUs or RAM memory usage). For such extended systems, accurate RCWA calculations are typically unfeasible while Finite Difference Time Domain calculations are computationally intensive. The 20 by 20 supercell systems were calculated utilizing the same amount of plane waves considered in the 4 by 4 benchmark configurations of the previous section. With the GMoUB, the only bottle-neck is that the Fourier transform operations need to be carried out over the extended domain. However, with available Fast Fourier Transform algorithms, the relevant operations are already memory and computationally efficient. Thus, the GMoUB could essentially allow one to address significantly larger areas so long one can ensure accurate Fourier Transform calculations. 

Here we consider the same layer stack and grating material as in the benchmarking calculations. 
\begin{figure}[h]

	\includegraphics[scale=0.27]{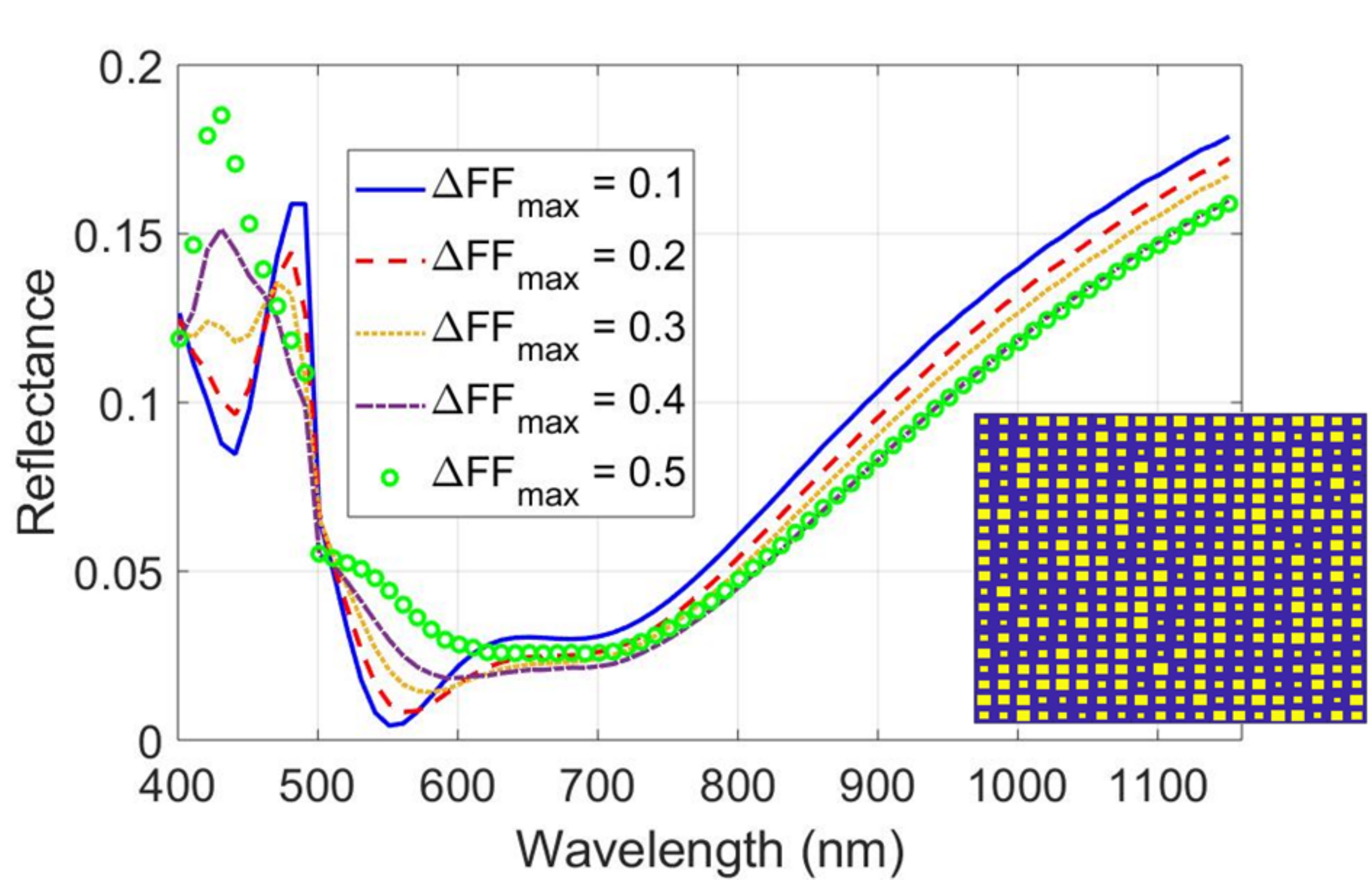}
	\caption{\textit{Reflectance of a  20 by 20 TiO$_2$ binary grating supercell system ontop of a high index 20~nm TiO$_2$/5~nm AlO$_\mathrm{x}$/ c-Si substrate multilayer stack with individual period  $P=500$~nm. The inset gives the supercell for $\Delta FF = 0.5$}} \label{fig7}

\end{figure}
We plot the reflectance of 20 by 20 grating supercells with varying unit cell fill factor perturbation for normal light incidence in Fig.~\ref{fig7}. An example of the realization of such supercells for the largest perturbation considered is sketched in the inset. Figure~\ref{fig7} shows that disorder can enhance light incoupling (reduction of reflectance) in the longer wavelength range $>$ 600~nm. The considered perturbed grating systems are coupling incoming light to the larger numbers of scattering pathways in the high index c-Si substrate in that range. Note however, that the shorter wavelength range exhibits instead an increase of reflectance. In this instance, the effect of opening more diffraction pathways due to disorder causes more light to be scattered back to air. As observed by many authors in the past\citep{Oskooi2012,Martins2013,Nanz2018,Piechulla2018}, we also demonstrate here that not all disorders are equal, and that increasing the disorder level may prevent the system from achieving the desired scattering response. 

Beyond the shallow perturbation regime, as we are considering now, a straightforward physical intuition may not be easily obtained, especially when considering disordered systems. We argue therefore that our method, which allows one to deduce in a fast and computationally efficient manner important trends caused by introducing disorder, would be of high importance in optimizing and analyzing such systems. Even more strongly perturbed systems can be considered if a reference different from our simplified unperturbed system is utilized. 

\section{summary}

We introduced the Green's method of unperturbed Bloch modes (GMoUB) for cost efficient forward modelling of disordered binary surface textures. A key approximation discussed here is the usage of Bloch modes of an unperturbed reference ordered system as ansatz in calculating the scattering response of disordered configurations. This approximation allows one to greatly reduce computational costs without sacrificing the ability to deduce the scattering response to all possible directions allowed by the disordered structure of interest beyond the shallow surface amplitude perturbation regime. We showed benchmarking calculations of our method against RCWA calculations with excellent agreement between both methods over an important regime of grating parameters and disorder of a physically relevant system for c-Si solar cell applications. As an example of our method's strength, we examined disordered decoupled binary light trapping textures for c-Si cells and demonstrated how the disorder may enhance light incoupling at a front solar cell interface. 

The GMoUB allows fast estimates of the scattering characteristics of spatially extended disordered textures beyond the shallow perturbation regime with small computational effort. The memory costs due to matrix inversion will be limited to the amount of plane waves we choose to represent the Bloch modes. Much like, RCWA, GMoUB allows one to go from truly rigorous to approximate simply by choosing the Bloch mode ansatz and the amount of plane waves representing each Bloch mode. The main advantage is the possibility to utilize a small amount of plane waves to represent the Bloch modes inside the scattering region, while still allowing one to estimate scattering response of all channels accessible by the disordered system.

\section{Acknowledgements}

This work has been funded by the German Federal Ministry of Education and Research, under the project SolarNano (grant no 13N13163) and the DFG Schwerpunktprogramm SPP1839 'Tailored Disorder' (RO 3640/6-1) and S\~{a}o Paulo Research Foundation (FAPESP) FAPESP/OSU 2015-AWARD 2015/50268-5.

\section{Appendix}

\begin{figure}[h!]

	\includegraphics[scale=0.25]{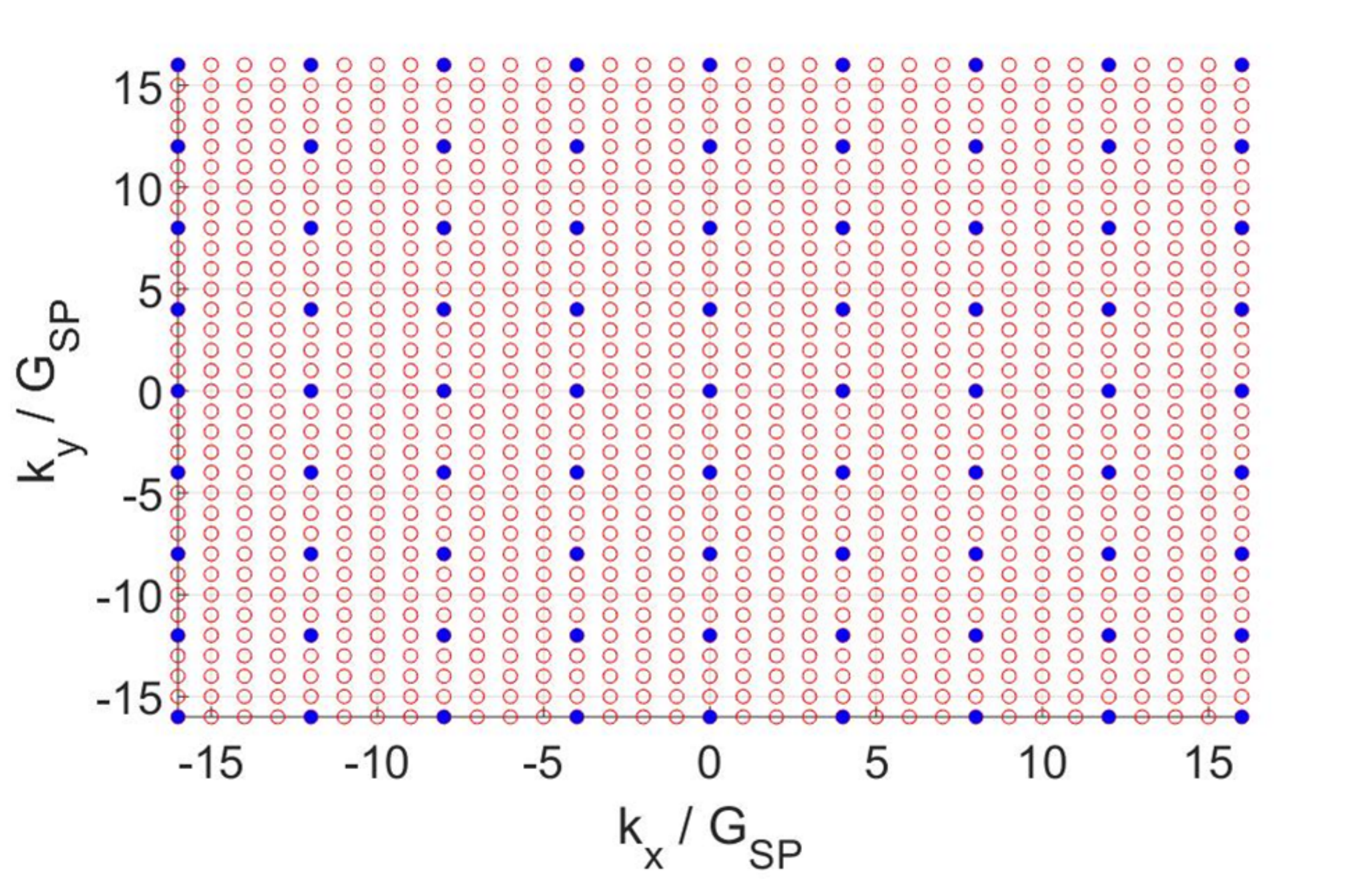}
	\caption{\textit{Sketch of the k-space mesh, which depicts plane components that are taken into account in the benchmarking calculations in section IV. The axes are k-vectors normalized to the supercell grating vector $G_\mathrm{SP}=2\pi/(4P)$. }} \label{fig8}

\end{figure}

We show a comparison of the k-space meshing between the RCWA and GMoUB calculation in Fig.~\ref{fig8}. Our GMoUB formalism considers plane waves indicated only by the blue filled circles in calculating the field in the scattering region (region II). Even though we consider such sparse k-space mesh in our GMoUB calculations, scattering in region I and III pertaining to the channels indicated by the red unfilled circles can still be deduced as discussed in section III and shown in Fig.~\ref{fig4}. RCWA calculations, on the other hand, would need to take into account all points, in order to deduce the scattering response to all those channels.

\begin{figure}[h!]

	\includegraphics[scale=0.16]{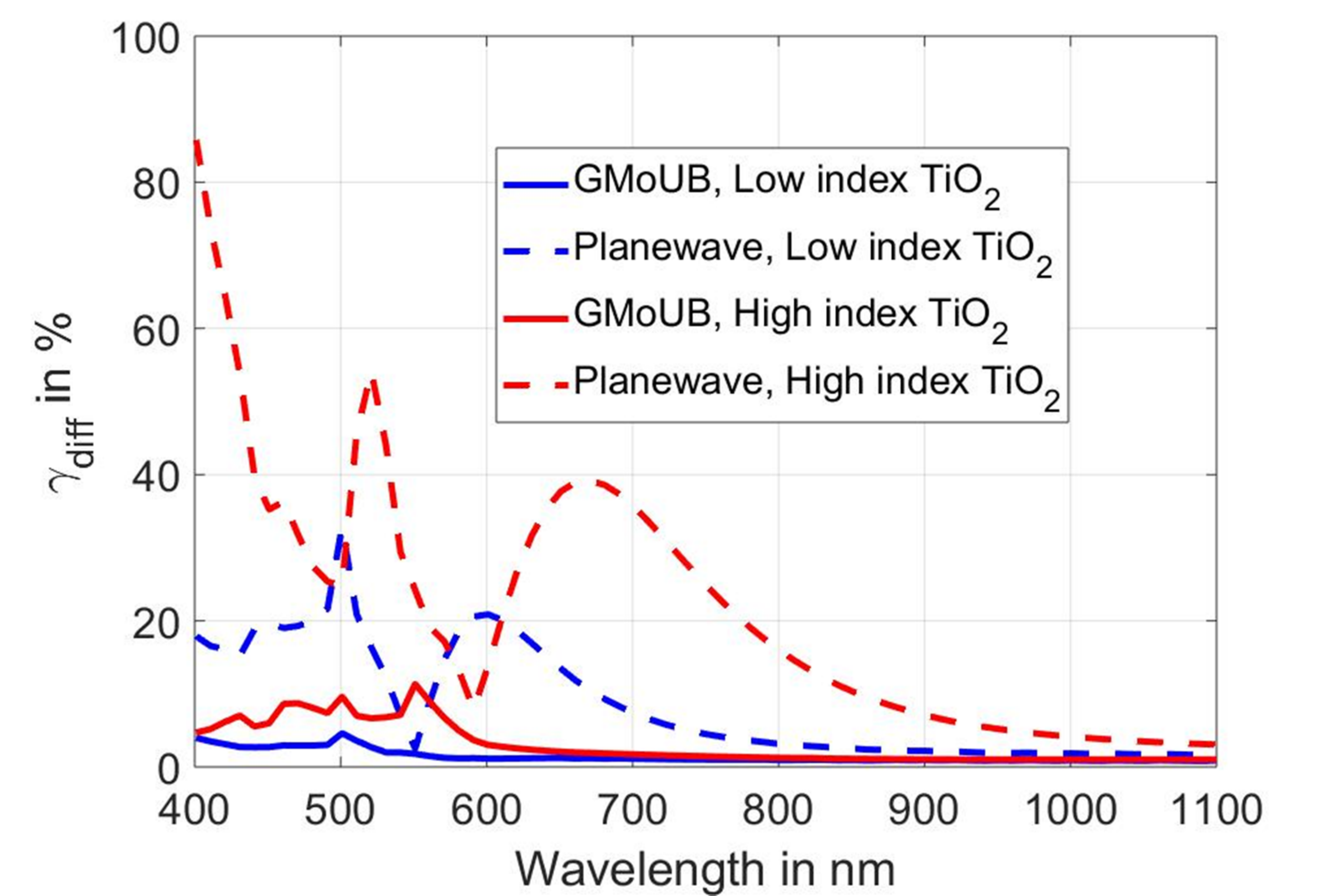}
	\caption{\textit{ Comparison of $\gamma_\mathrm{diff}$ for $\Delta FF=0.05$ and $h_0=100$~nm of the 4 by 4 grating structure described in Fig.~\ref{fig3} when calculated assuming plane waves (Eq.~\ref{eq:Flatmodes-1}) in region I or Bloch modes (Eq.~\ref{eq:Bloch}) of a perfectly ordered system ($FF=0.5$) as ansatz for the field in region II. All the calculations displayed here are done utilizing the same k-space meshing as described in section IV for the GMoUB.}} \label{fig9}

\end{figure}

An error analysis comparing the usage of Eq.~\ref{eq:Flatmodes-1} and Eq.~\ref{eq:Bloch} to considering the benchmarking structures are given in Fig.~\ref{fig9}. The plane wave ansatz displays fairly large error  although there is minimum perturbation in the fill factor, especially at shorter wavelengths.

\bibliography{references2}

\end{document}